\documentclass[corporate, usegrafix]{adqgem}

\font\myf=cmssq8 at 4pt

\newenvironment{narrow}[2]{%
  \begin{list}{}{%
    \setlength{\topsep}{10pt}%
    \setlength{\leftmargin}{#1}%
    \setlength{\rightmargin}{#2}%
    \setlength{\listparindent}{\parindent}%
    \setlength{\itemindent}{\parindent}%
    \setlength{\parsep}{\parskip}}%
    \item[]}{\end{list}}

\newcommand{\scalt}{\scal{t}}
\newcommand{\scali}{\scal{i}}
\newcommand{\scalm}{\scal{m}}
\newcommand{\scalh}{\scal{h}}
\newcommand{\scalG}{\scal{G}}
\newcommand{\scalQ}{\scal{Q}}
\newcommand{\scalR}{\scal{R}}
\newcommand{\scalS}{\scal{S}}
\newcommand{\scalV}{\scal{V}}
\newcommand{\scaldQ}{d\scal{Q}}
\newcommand{\scaldi}{d\scal{i}}

\newcommand{\vectr}{\vect{r}}
\newcommand{\vecta}{\vect{a}}
\newcommand{\vectu}{\vect{u}}
\newcommand{\vectj}{\vect{j}}
\newcommand{\vectg}{\vect{g}}
\newcommand{\vectb}{\vect{b}}
\newcommand{\vects}{\vect{s}}
\newcommand{\vectD}{\vect{D}}
\newcommand{\vectH}{\vect{H}}
\newcommand{\vectB}{\vect{B}}
\newcommand{\vectE}{\vect{E}}
\newcommand{\vectN}{\vect{N}}

\newcommand{\vectdS}{d\vect{S}}

\newcommand{\ke}{k_e}
\newcommand{\km}{k_m}
\newcommand{\kg}{k_g}
\newcommand{\dQdV}{dQ/dV}
\newcommand{\didV}{di/dV}

\newcommand{\vectJE}{\vect{J}_e}
\newcommand{\vectJM}{\vect{J}_m}
\newcommand{\vectJG}{\vect{J}_g}
\newcommand{\vectkap}{{\mathbf\kappa}}

\newcommand{\zzz}{\alpha}
\newcommand{\mz}{\mu\zzz}
\newcommand{\emz}{\zzz\epsilon\mu}

\newcommand{\induces}{$\rightarrow$}
\newcommand{\brackD}{$\left<\vectH, \vectD\right>$}
\newcommand{\brackB}{$\left<\vectE, \vectB\right>$}
\newcommand{\brackg}{$\left<\vectN, \vectg\right>$}

\newcommand{\nbox}{{$\mathbf\bullet$}}
\newcommand{\embox}{{$\mathbf\pi$}}
\newcommand{\gebox}{{$\mathbf\kappa$}}
\newcommand{\gmbox}{{$\mathbf\lambda$}}
\newcommand{\gembox}{{$\mathbf\omega$}}

\newcommand{\multibox}[4]{{#1#2#3#4}}
\newcommand{\xnnnn}{\multibox{\nbox}{\nbox}{\nbox}{\nbox}}
\newcommand{\xannn}{\multibox{\embox}{\nbox}{\nbox}{\nbox}}
\newcommand{\xbnnn}{\multibox{\nbox}{\gebox}{\nbox}{\nbox}}
\newcommand{\xcnnn}{\multibox{\nbox}{\nbox}{\gmbox}{\nbox}}
\newcommand{\xdnnn}{\multibox{\nbox}{\nbox}{\nbox}{\gembox}}

\newcommand{\xadnn}{\multibox{\embox}{\nbox}{\nbox}{\gembox}}
\newcommand{\xbdnn}{\multibox{\nbox}{\gebox}{\nbox}{\gembox}}
\newcommand{\xcdnn}{\multibox{\nbox}{\nbox}{\gmbox}{\gembox}}

\newcommand{\xabdn}{\multibox{\embox}{\gebox}{\nbox}{\gembox}}
\newcommand{\xacdn}{\multibox{\embox}{\nbox}{\gmbox}{\gembox}}
\newcommand{\xbcdn}{\multibox{\nbox}{\gebox}{\gmbox}{\gembox}}
\newcommand{\xabcd}{\multibox{\embox}{\gebox}{\gmbox}{\gembox}}

\newcommand{\signat}[3]{{H#1~-~E#2~-~N#3}}
\newcommand{\ctr}[1]{\multicolumn{1}{|c|}{#1}}
\newcommand{\sig}[4]{{\signat{#1}{#2}{#3}\;\tiny\hfill #4}}

\begin{document}

\workno{GEM}{2000}{01D}
\worker{A. I. A. Adewole}
\monthdate{April 2000}
\mailto{aiaa@adequest.ca}
\work{Unification By Induction}
\shortwork{Unification By Induction}
\makefront

\begin{abstract}
We show that the problem of unifying electromagnetism with gravity
has an elegant solution in classical physics through the phenomenon
of induction. By studying the way that induction leads to the formation 
of electromagnetic fields, we identify the classical field
equations which the unified field must satisfy and a
corresponding set of constitutive equations for the medium
sustaining the field. The unification problem is then reduced to
the problem of finding the exact form of these constitutive
equations for different media by experiments.
\end{abstract}

\section{Introduction}\label{INTRO}
The problem of unifying gravitational fields with electric and
magnetic fields is known to be one of the most pressing in
contemporary physics, not only for theoretical reasons, but also
because of the many uses to which electromagnetic fields which
are formed by the unification of electric and magnetic fields
have been put in science, engineering and business. 
While many different solutions to the problem have been
proposed~\cite{Green99, Maartens98, Magnon92, Schwarz92}, 
to date none of these solutions has had much success 
in the sense of having the character of the unification accomplished 
by Maxwell. One reason why Maxwell may have
succeeded where current theories fail is that he was guided by a
mechanism describing the formation of an electromagnetic field
while current theories abolish any such mechanism. By this we mean that an
electromagnetic field is treated nowadays as a
monolithic entity that manifests sometimes as a magnetic field
and sometimes as an electric field, and not as a more complex
field formed by interaction between the much simpler electric and
magnetic fields. A consequence of this treatment is that one has no 
extra degree of freedom or guiding principle with which to introduce 
gravity into the framework of Maxwell's theory. 
On the other hand, if we grant for a moment that an electromagnetic field 
is a composite entity, it follows that we must first understand the
process by which it is formed before tackling the problem of
unifying it with gravity. In this case the problem of unification will
have a different meaning and scope, namely, to find the process by
which electric, magnetic and gravitational fields can interact
with one another to form more complex (electromagnetic,
electrogravitic, etc) fields in a medium. 

In Maxwell's early approach to his theory, this process was assumed to be
mechanical and an attempt was made to construct a detailed model
of its operations based on different ether theories of the time.
In more recent theories, attempts are made to construct a model by
which gravity may be reduced to electromagnetism~\cite{Assis92, Assis95}. 
While these model-building approaches are not without merit, 
they appear to be inherently superficial, and for that reason, we 
shall focus on obtaining a solution to the problem from 
a different approach. We elaborate this new approach, which is based on 
Newton's theory of gravitation and Hertz's version~\cite{Hertz62} of 
Maxwell's theory~\cite{Maxwell54}, by first giving
an interpretation of Hertz's equations that
eliminates the ether concept as in ~\cite{Phipps93,
Phipps86} but has the advantage that in it, the equations are not
postulated but derived from first principles. The derivation will
show that Hertz's equations differ from Maxwell's equations due
to the way the concepts of electric current and emf are treated
in Hertz's theory. This will be shown in the next two sections.
In \secref{SSEM} we will describe how electric and
magnetic fields interact through induction to form an
electromagnetic field in Hertz's theory. In \secref{SSG}
and \secref{SSGEM} we will show that induction reduces the
problem of unifying electromagnetism and gravity to the problem
of finding the constitutive equations for the unified field if
Newton's theory is taken into account. In \secref{SSPMG}
and \secref{SSCGM} we will discuss some implications of
this method of unification. Finally, in 
\secref{SSDAC}, we will conclude the paper with a discussion of
the main results and present some suggestions for further
research\footnote{It is pertinent to note here that ``Maxwell's theory'' 
as used in the rest of this paper refers to the theory as currently 
understood and not as originally proposed by Maxwell.}.

\section{Electric Fields}\label{SSE}
We begin by assuming that in Hertz's theory, the strength or
intensity of an electric field can be characterized at any point
\vectr\ and time \scalt\ by a vector which we shall denote by
\vectD(\vectr, \scalt). Since it will clearly not be possible for us to
describe an electric field mathematically in Hertz's theory if
this vector is all we have, we shall introduce other parameters
in addition to this vector. But in order for us to be able to
compare Hertz's theory with Maxwell's, these additional
parameters must be chosen to correspond as closely as possible to
familiar quantities. For this reason we introduce parameter
\scalQ\ which is the flux of \vectD\ through an element \vectdS\
of a surface \scalS:
\begin{equation}\label{E:dQ}
\scaldQ = \dprod{\vectD}{\vectdS}
\end{equation}
and parameter \scali\ which is the flux of the time derivative of
\vectD\ through the surface element \vectdS:
\begin{equation}\label{E:di}
\scaldi = \dprod{\vectJE}{\vectdS}, \quad
\vectJE = d\vectD/d\scalt.
\end{equation}
In familiar terms parameter \scalQ\ may be called an electric
charge while parameter \scali\ may be called an electric current,
and since these equations are the definitions of these parameters
in Hertz's theory, vector \vectD\ is fundamental and may be
regarded as the source of \scalQ\ while its time derivative
$\vectJE$ may be regarded as the source of \scali. This means
that unlike in Maxwell's theory where charges and currents give
cause to a field, in Hertz's theory, it is the field that gives
cause to charge and current.

Now if \scalS\ is a closed surface bounding a region \scalR\ of
space, and if Gauss' divergence theorem~(\cite{Tromba88}, p532)
holds for
\vectD\ in this region, it follows from \eqnref{E:dQ} that
\begin{equation}\label{E:divD}
\dive{\vectD} = \sigma
\end{equation}
where $\sigma = \dQdV$ and $dV$ is a volume element of \scalR.
Similarly, if Gauss' theorem holds for $\vectJE$ in \scalR, it
follows from \eqnref{E:di} that
\begin{equation}\label{E:divJE}
\dive{\vectJE} = \eta
\end{equation}
where $\eta = \didV$. When the electric current \scali\ in a
region is independent of the volume \scalV\ of the region, as
experiments show to be the case (e.g. in thin and thick wires), we
have $\eta = 0$ and the above equation implies that a vector
\vectH\ exists such that
\begin{subequations}\label{E:curlH}
\begin{equation}\label{E:curlHa}
\curl{\vectH} = \vectJE.
\end{equation}
This is Ampere's law. If experiments were to show that the
electric current \scali\ in a region varies linearly with the
volume \scalV\ of the region, then $\eta = 3\ke =
\text{constant}$ and it will follow from \eqnref{E:divJE} that
\begin{equation}\label{E:curlHb}
\curl{\vectH} = \vectJE - \ke\vectr.
\end{equation}
\end{subequations}
Thus Ampere's law may have different forms for different $\eta$
in Hertz's theory.

Inspection shows that there is nothing in the derivation of
\eqnref{E:curlH} to suggest that \vectH\ describes a field in the
physical sense. If we assume that \vectH\ does describe a field,
there is nothing in the derivation to suggest that the field
described by \vectH\ is physically caused or induced by \vectD.
If we assume this as well, there is still nothing in the
derivation to suggest that the field described by
\vectH\ is magnetic. As we shall see later, {\em one cannot
always assume that \vectH\ describes a magnetic field in Hertz's
theory.} Differentiating \vectD(\vectr, \scalt) with respect to
time gives
\begin{equation}\label{E:JE}
\vectJE = \partial\vectD/\partial\scalt + \vectj, \quad
\vectj = \udiv{\vectD}
\end{equation}
where $\vectu = d\vectr/dt$. Vector identities for the curl of a
vector product and the gradient of a scalar
product~(\cite{Aris89}, p57) can be used to rewrite \vectj\ as
\begin{subequations}\label{E:j}
\begin{align}\label{E:ja}
\vectj &= \sigma\vectu + \curl{(\cprod{\vectD}{\vectu})} -
\vectD(\dive{\vectu}) + (\dprod{\vectD}{\nabla})\vectu
\\
\label{E:jb}
\vectj &= \grad{(\dprod{\vectD}{\vectu})} -
(\dprod{\vectD}{\nabla})\vectu -
\cprod{\vectD}{(\curl{\vectu})} -
\cprod{\vectu}{(\curl{\vectD})}
\end{align}
\end{subequations}
which shows that Hertz's theory has a convection current density
$\sigma\vectu$ and a conduction current density $\vectj -
\sigma\vectu$ like Maxwell's theory. The significance of this
result will be discussed later.

\section{Magnetic Fields}\label{SSM}
By analogy with electric fields, we assume that in Hertz's theory
the intensity of a magnetic field can be characterized at point
\vectr\ and time \scalt\ by a vector \vectB(\vectr, \scalt). To
describe the field mathematically, we introduce parameter $\Phi$
which is the flux of \vectB\ through an element \vectdS\ of a 
surface \scalS\:
\begin{equation}\label{M:dPhi}
d\Phi = \dprod{\vectB}{\vectdS}
\end{equation}
and parameter $\xi$ which is the inward flux of the time
derivative of \vectB\ through \vectdS:
\begin{equation}\label{M:dxi}
d\xi = \dprod{\vectJM}{\vectdS}, \quad
\vectJM = -d\vectB/d\scalt.
\end{equation}
In familiar terms parameter $\Phi$ may be called a magnetic flux while
parameter $\xi$ may be called an emf, but in Hertz's theory, one may also
call $\Phi$ a magnetic charge and $\xi$ a magnetic current by
analogy with \eqnref{E:dQ} and \eqnref{E:di}.

If \scalS\ is a closed surface bounding a region \scalR\ of
space, and if Gauss' theorem holds for \vectB\ in this region, it
follows from \eqnref{M:dPhi} that
\begin{equation}\label{M:divB}
\dive{\vectB} = \delta
\end{equation}
where $\delta = d\Phi/dV$ and $dV$ is a volume element of \scalR.
Similarly, if Gauss' theorem holds for $\vectJM$ in \scalR, it
follows from \eqnref{M:dxi} that
\begin{equation}\label{M:divJM}
\dive{\vectJM} = \vartheta
\end{equation}
where $\vartheta = d\xi/dV$. In particular when the emf $\xi$ in
a region is independent of the volume \scalV\ of the region, as
experiments show to be the case (e.g. in small and large voltaic
cells), we have $\vartheta = 0$ and the last equation implies that
a vector \vectE\ exists such that
\begin{subequations}\label{M:curlE}
\begin{equation}\label{M:curlEa}
\curl{\vectE} = \vectJM.
\end{equation}
This is Faraday's law. If experiments were to show that $\xi$
varies linearly with the volume \scalV\ of the region, then
$\vartheta = 3\km = \text{constant}$ and it will follow from
\eqnref{M:divJM} that
\begin{equation}\label{M:curlEb}
\curl{\vectE} = \vectJM - \km\vectr.
\end{equation}
\end{subequations}
Thus Faraday's law may have different forms for different
$\vartheta$ in Hertz's theory.

Equation \eqnref{M:curlEa} differs from Faraday's law in Maxwell's
theory in two ways. First, there is nothing in the derivation of
this equation to suggest that \vectE\ describes a field or that
the field it describes is electric or physically induced by
\vectB. We may assume that \vectE\ describes a field induced by
\vectB, but as we shall see later, {\em one cannot always assume that
\vectE\ describes an electric field in Hertz's theory}. Second,
this equation contains an extra term that is not present in
Maxwell's theory. We can see this clearly by differentiating
\vectB(\vectr, \scalt) with respect to time to get
\begin{equation}\label{M:JM}
\vectJM = -\partial\vectB/\partial\scalt - \vectb, \quad
\vectb = \udiv{\vectB}
\end{equation}
which, when substituted into \eqnref{M:curlEa}, yields
\begin{equation}\label{M:curlEc}
\curl{\vectE} = -\partial\vectB/\partial\scalt - \vectb.
\end{equation}
The term \vectb\ in this equation arises because of the way an
emf $\xi$ is defined in Hertz's theory. For if we were to define
an emf $\xi^\prime$ by
\begin{equation}\label{M:mwl1}
\xi^\prime = -d\Phi/dt,
\end{equation}
as in Maxwell's theory, then one can take \eqnref{M:dPhi} into
account to get
\begin{equation}\label{M:mwl2}
\xi^\prime = -\frac{d}{dt}\int\dprod{\vectB}{\vectdS}
\end{equation}
so that by Reynold's transport theorem~(\cite{Aris89}, p85),
\begin{equation}\label{M:mwl3}
\xi^\prime = -\int\dprod{(\partial\vectB/\partial t +
\delta\vectu)}{\vectdS}
\end{equation}
and by Gauss' theorem,
\begin{equation}\label{M:mwl4}
d\xi^\prime/dV = -\dive{(\partial\vectB/\partial t +
\delta\vectu)}.
\end{equation}
Thus when $d\xi^\prime/dV = 0$, we get $\curl{\vectE} = -
\partial\vectB/\partial t - \delta\vectu$ and since $\delta = 0$
identically in Maxwell's theory, Faraday's law is obtained as
$\curl{\vectE} = -\partial\vectB/\partial t$.

It is interesting to note that if one defines an electric current
not by \eqnref{E:di} but as
\begin{equation}\label{M:mwl5}
i^\prime = dQ/dt,
\end{equation}
by analogy with \eqnref{M:mwl1}, then by \eqnref{E:dQ} one would have
\begin{equation}\label{M:mwl6}
i^\prime = \frac{d}{dt}\int\dprod{\vectD}{\vectdS}
\end{equation}
so that by Reynold's theorem,
\begin{equation}\label{M:mwl7}
i^\prime = \int\dprod{(\partial\vectD/\partial t +
\sigma\vectu)}{\vectdS}
\end{equation}
and by Gauss' theorem,
\begin{equation}\label{M:mwl8}
di^\prime/dV = \dive{(\partial\vectD/\partial t + \sigma\vectu)}.
\end{equation}
When $di^\prime/dV = 0$, one gets Ampere's law as
\begin{equation}\label{M:mwl9}
\curl{\vectH} = \partial\vectD/\partial t + \vectj, \quad
\vectj = \sigma\vectu
\end{equation}
which shows that if \eqnref{M:mwl5} is correct, then a conduction
current density $\vectj - \sigma\vectu$ should not exist. 

But if
one were to define an electric current by \eqnref{E:di} as in Hertz's
theory, then by subtracting \eqnref{M:mwl7} from \eqnref{E:di} and
taking \eqnref{E:JE} into account, one gets
\begin{equation}\label{M:ij}
i = dQ/dt + \int\dprod{(\vectj - \sigma\vectu)}{\vectdS}
\end{equation}
instead of \eqnref{M:mwl5}. Similarly, by subtracting \eqnref{M:mwl3}
from \eqnref{M:dxi} and taking \eqnref{M:JM} into account, one gets
\begin{equation}\label{M:xib}
\xi = -{d\Phi/dt} - \int\dprod{(\vectb - \delta\vectu)}{\vectdS}
\end{equation}
instead of \eqnref{M:mwl1}. To calculate $\xi$ by this equation it is
useful to rewrite \vectb\ as
\begin{subequations}\label{M:b}
\begin{align}\label{M:ba}
\vectb &= \delta\vectu + \curl{(\cprod{\vectB}{\vectu})} -
\vectB(\dive{\vectu}) + (\dprod{\vectB}{\nabla})\vectu
\\
\label{M:bb}
\vectb &= \grad{(\dprod{\vectB}{\vectu})} -
(\dprod{\vectB}{\nabla})\vectu -
\cprod{\vectB}{(\curl{\vectu})} -
\cprod{\vectu}{(\curl{\vectB})}
\end{align}
\end{subequations}
by analogy with \eqnref{E:j}.

\section{Formation Of Electromagnetic Fields}\label{SSEM}
We have noted that in Hertz's theory, one can only
assume that \vectE\ describes a field induced by \vectB\ because
there is nothing in the derivation of \eqnref{M:curlE} to suggest
that the field described by \vectE\ is electric. If it is known
however that \vectB\ induces an electric field in a particular
medium, then one may for this reason assume that \vectE\
describes an electric field in that medium. Theoretically, the
possibility must be allowed that \vectB\ may induce different
kinds of fields in different media. For example, as far as theory
is concerned, there may be media in which \vectB\ induces a
magnetic field. In such media one must assume that \vectE\
describes a magnetic field according to Hertz's theory.
Furthermore, if \vectE\ describes a magnetic field or an electric
field in a medium, then according to Hertz's theory, one should
expect \vectE\ to be a function of \vectB\ or \vectD\ in the
medium respectively.
To illustrate this idea (which we shall call the principle of
constitution), suppose we know that \vectD\ induces a magnetic field
and that \vectB\ induces an electric field in a medium. Then in
this medium, \vectH\ (i.e. the field induced by \vectD) describes
a magnetic field and is therefore a function of \vectB\ while
\vectE\ (i.e. the field induced by \vectB) describes an electric
field and is therefore a function of \vectD:
\begin{subequations}\label{EM:con}
\begin{equation}\label{EM:cona}
\vectH = \vectH(\vectB),\quad
\vectE = \vectE(\vectD).
\end{equation}
What if \vectD\ induces an electric field and \vectB\ induces a
magnetic field in a medium ? Then in this medium,
\begin{equation}\label{EM:conb}
\vectH = \vectH(\vectD),\quad
\vectE = \vectE(\vectB).
\end{equation}
What if \vectD\ induces both types of fields in a medium and
\vectB\ does likewise ? Then in this medium,
\begin{equation}\label{EM:conc}
\vectH = \vectH(\vectD, \vectB),\quad
\vectE = \vectE(\vectD, \vectB).
\end{equation}
\end{subequations}
Constitutive equations like these describe interactions between
the electric field \vectD\ and the magnetic field \vectB\ in
Hertz's theory in the following sense.

Suppose that \eqnref{EM:cona} holds in a medium. 
Then in this medium \vectB\ induces a field
\vectE\ that is related to \vectD\ while \vectD\ induces a field 
\vectH\ that is related to \vectB. Fields \vectB\ and \vectD\ interact 
in this way with each other to form an electromagnetic field in the
medium. In the same way if \eqnref{EM:conc} holds in a medium,
each field interacts with itself as well as with the other field
so that one can speak of the presence of an electromagnetic field
in the medium. But if \eqnref{EM:conb} holds in a medium, \vectD\
induces a field \vectH\ that is related to \vectD\ while \vectB\ 
induces a field \vectE\ that is related to \vectB. Each of \vectB\ 
and \vectD\ interacts with itself but not with the other field and they
therefore do not form an electromagnetic field in the medium.
These examples give a clear picture of how an electromagnetic
field is formed in Hertz's theory. One sees clearly that
\eqnref{EM:conb} is not admissible as constitutive equations for an
electromagnetic field in Hertz's theory. This fact is taken for 
granted in Maxwell's theory where the formation of an electromagnetic
field is less explicit. The fundamental vectors
of an electromagnetic field in Hertz's theory are also seen to be
\vectD\ and \vectB, not \vectE\ or \vectH, because according to
the principle of constitution {\em the nature of the fields
described by \vectE\ and \vectH\ is not known until the
constitutive equations for a medium are given}, whereas \vectD\
always describes an electric field and \vectB\ always describes a
magnetic field. Thus Hertz's theory leaves no room for debate
concerning the physical interpretation of these 
vectors~\cite{Roche00, Chambers99}.

\section{Unification With Gravity}\label{SSG}
Electromagnetism and gravity can be unified by the constitution principle
if we assume that electric and magnetic fields can
induce a gravitational field and that a gravitational field can
induce electric and magnetic fields. We need two more vectors in
order to perform the unification theoretically: one vector
\vectg\ to describe a gravitational field and another vector
\vectN\ to describe a field induced by \vectg. Using these
vectors, interactions between electric, magnetic and
gravitational fields can be described as in the previous section.
Thus if \vectB\ induces a gravitational field, then \vectE\
describes a gravitational field and is therefore a function of
\vectg. If \vectg\ induces an electric field, then \vectN\
describes an electric field and is therefore a function of
\vectD, et cetera. Similarly to how an electromagnetic field is
formed, a gravoelectromagnetic field may be formed in a medium if
electric, magnetic and gravitational fields interact mutually in
the medium. This may occur in the simplest case if \vectg\
induces an electric field, \vectD\ induces a magnetic field, and
\vectB\ induces a gravitational field in a medium:
\begin{subequations}\label{G:con}
\begin{equation}\label{G:cona}
\vectN = \vectN(\vectD), \quad
\vectH = \vectH(\vectB), \quad
\vectE = \vectE(\vectg)
\end{equation}
or alternatively, if \vectg\ induces a magnetic field, \vectB\
induces an electric field, and \vectD\ induces a gravitational
field in the medium:
\begin{equation}\label{G:conb}
\vectN = \vectN(\vectB), \quad
\vectE = \vectE(\vectD), \quad
\vectH = \vectH(\vectg).
\end{equation}
\end{subequations}
As we shall see later, these equations are a special case of a
more general set of equations.

We need a suitable gravitational theory for \vectg\ and \vectN\
in order to complete the unification. If Newton's theory is
adopted for this purpose, the strength or intensity of a
gravitational field can be characterized at any point \vectr\ and
time \scalt\ by a vector $\vectg(\vectr, \scalt) =
\vecta/4\pi\scalG$ where \vecta(\vectr, \scalt) is the
acceleration due to gravity and \scalG\ is the gravitational
constant\footnote{We observe here that \vectg\ should be regarded 
as a fundamental quantity while $\vectg =\vecta/4\pi\scalG$ should be 
regarded as a hypothetical relation and not as a definition of \vectg.}. 
We introduce parameter \scalm\ which is the flux of
\vectg\ through an element \vectdS\ of a surface \scalS:
\begin{equation}\label{G:dm}
d\scalm = \dprod{\vectg}{\vectdS}
\end{equation}
and parameter \scalh\ which is the flux of the time derivative of
\vectg\ through \vectdS:
\begin{equation}\label{G:dh}
d\scalh = \dprod{\vectJG}{\vectdS}, \quad
\vectJG = d\vectg/d\scalt.
\end{equation}
In familiar terms parameter \scalm\ is clearly a gravitational
mass, but as there is no obvious name for parameter \scalh, let
us call it a gravitational current for the moment.

If \scalS\ is a closed surface bounding a region \scalR\ of
space, and if Gauss' theorem holds for \vectg\ in this region,
then from \eqnref{G:dm} we have
\begin{equation}\label{G:divg}
\dive{\vectg} = \rho
\end{equation}
where $\rho = d\scalm/dV$ and $dV$ is a volume element of \scalR.
Similarly, if Gauss' theorem holds for $\vectJG$ in this region,
then from \eqnref{G:dh} we have
\begin{equation}\label{G:divJG}
\dive{\vectJG} = \varpi
\end{equation}
where $\varpi = d\scalh/dV$. In particular if \scalh\ does not
depend on the volume \scalV\ of a region, then $\varpi = 0$ and it
follows from the last equation that a vector \vectN\ exists such
that
\begin{subequations}\label{G:curlN}
\begin{equation}\label{G:curlNa}
\curl{\vectN} = \vectJG.
\end{equation}
This is the gravitational analogue of Ampere's and Faraday's laws.
If \scalh\ were to vary linearly with the volume \scalV\ of a
region, then $\varpi = 3\kg = \text{constant}$ and it will follow
from \eqnref{G:divJG} that
\begin{equation}\label{G:curlNb}
\curl{\vectN} = \vectJG - \kg\vectr.
\end{equation}
\end{subequations}
Differentiating \vectg(\vectr, \scalt) gives
\begin{equation}\label{G:JG}
\vectJG = \partial\vectg/\partial\scalt + \vects, \quad
\vects = \udiv{\vectg}
\end{equation}
while similarly to \eqnref{E:j} we have
\begin{subequations}\label{G:s}
\begin{align}\label{G:sa}
\vects &= \rho\vectu + \curl{(\cprod{\vectg}{\vectu})} -
\vectg(\dive{\vectu}) + (\dprod{\vectg}{\nabla})\vectu
\\
\label{G:sb}
\vects &= \grad{(\dprod{\vectg}{\vectu})} -
(\dprod{\vectg}{\nabla})\vectu -
\cprod{\vectg}{(\curl{\vectu})} -
\cprod{\vectu}{(\curl{\vectg})}
\end{align}
\end{subequations}
and by arguing as in the derivation of \eqnref{M:ij} we get
\begin{equation}\label{G:hs}
h = dm/dt + \int\dprod{(\vects - \rho\vectu)}{\vectdS}.
\end{equation}
These equations show that a gravitational current is a mass flow,
so there is a familiar name for the concept after all.

\section{Gravoelectromagnetic Waves}\label{SSGEM}
We have synthesized Newton's theory and Hertz's version of
Maxwell's theory into a gravoelectromagnetic (GEM) theory that
unifies electromagnetism and gravity in classical terms. Our
results show that an electric field \vectD\ is described by the
field equations
\begin{subequations}\label{GEM:E}
\begin{equation}\label{GEM:Ea}
\dive{\vectD} = \sigma
\end{equation}
\begin{equation}\label{GEM:Eb}
\curl{\vectH} = \partial\vectD/\partial\scalt + \vectj
\end{equation}
\begin{equation}\label{GEM:Ec}
\vectj = \udiv{\vectD}
\end{equation}
\end{subequations}
while a magnetic field \vectB\ is described by the field equations
\begin{subequations}\label{GEM:M}
\begin{equation}\label{GEM:Ma}
\dive{\vectB} = \delta
\end{equation}
\begin{equation}\label{GEM:Mb}
\curl{\vectE} = -\partial\vectB/\partial\scalt - \vectb
\end{equation}
\begin{equation}\label{GEM:Mc}
\vectb = \udiv{\vectB}
\end{equation}
\end{subequations}
and a gravitational field \vectg\ is described by the field
equations
\begin{subequations}\label{GEM:G}
\begin{equation}\label{GEM:Ga}
\dive{\vectg} = \rho
\end{equation}
\begin{equation}\label{GEM:Gb}
\curl{\vectN} = \partial\vectg/\partial\scalt + \vects
\end{equation}
\begin{equation}\label{GEM:Gc}
\vects = \udiv{\vectg}.
\end{equation}
\end{subequations}
The fields interact in a medium through induction and their
interaction is described mathematically by the constitutive
equations of the medium. Following the discussion in the last two
sections, it is clear that the most general type of interaction
occurs if \vectg\ induces all three types of fields in a medium,
\vectD\ does the same, and likewise \vectB:
\begin{equation}\label{GEM:con}
\vectN = \vectN(\vectD, \vectB, \vectg), \quad
\vectH = \vectH(\vectD, \vectB, \vectg), \quad
\vectE = \vectE(\vectD, \vectB, \vectg).
\end{equation}
This is the general set of equations for which \eqnref{G:con} and \eqnref{EM:con}
are special cases.

Now if GEM fields exist, it may be possible for them to propagate
through space as waves. As a trivial but illustrative example of
how this might occur, suppose that \eqnref{G:cona} holds in a medium
and that in this medium, the equation has the form
\begin{equation}\label{GEM:sim}
\vectD = \epsilon\vectN, \quad
\vectB = \mu\vectH, \quad
\vectg = \zzz\vectE
\end{equation}
where $\epsilon, \mu, \zzz$ are independent of time \scalt\ and
position \vectr. Then in this medium, the field equations will be
\begin{subequations}\label{GEM:simE}
\begin{equation}\label{GEM:simEa}
\dive{\vectD} = \sigma
\end{equation}
\begin{equation}\label{GEM:simEb}
\curl{\vectB} = \mu(\partial\vectD/\partial\scalt + \vectj)
\end{equation}
\begin{equation}\label{GEM:simEc}
\vectj = \udiv{\vectD}
\end{equation}
\end{subequations}
\begin{subequations}\label{GEM:simM}
\begin{equation}\label{GEM:simMa}
\dive{\vectB} = \delta
\end{equation}
\begin{equation}\label{GEM:simMb}
\curl{\vectg} = -\zzz(\partial\vectB/\partial\scalt + \vectb)
\end{equation}
\begin{equation}\label{GEM:simMc}
\vectb = \udiv{\vectB}
\end{equation}
\end{subequations}
\begin{subequations}\label{GEM:simG}
\begin{equation}\label{GEM:simGa}
\dive{\vectg} = \rho
\end{equation}
\begin{equation}\label{GEM:simGb}
\curl{\vectD} = \epsilon(\partial\vectg/\partial\scalt + \vects)
\end{equation}
\begin{equation}\label{GEM:simGc}
\vects = \udiv{\vectg}.
\end{equation}
\end{subequations}
Taking the curl of \eqnref{GEM:simMb} twice and using
\eqnref{GEM:simEb} and \eqnref{GEM:simGb} to eliminate $\curl{\vectB}$
and $\curl{\vectD}$ from the result gives
\begin{equation}\label{GEM:wav1}
\curl{\lapl{\vectg}} - \emz\partial^3\vectg/\partial\scalt^3 =
\zzz\curl{(\curl{\vectb})} + \emz\partial^2\vects/\partial\scalt^2
+ \mz\partial(\curl{\vectj})/\partial\scalt.
\end{equation}
The terms containing \vectb, \vects\ and \vectj\ can be
simplified if we know \vectu(\vectr, \scalt). Restricting
ourselves to uniform translational motion, for which \vectu\ is
independent of time and position, it is easy to verify that
the following equalities hold:
\newcommand{\lpd}{(\partial\vectg/\partial\scalt)}
\newcommand{\qpd}{\udiv{(\partial^2\vectg/\partial\scalt^2})}
\begin{align*}
\partial^2\vects/\partial\scalt^2 &= \qpd\\
\curl{(\curl{\vectb})} &= \epsilon\mu\qpd +
2\epsilon\mu\udivn{2}{\lpd} + \epsilon\mu\udivn{3}{\vectg}\\
\partial(\curl{\vectj})/\partial\scalt &= \epsilon\qpd +
\epsilon\udivn{2}{\lpd}.
\end{align*}
Substituting these into \eqnref{GEM:wav1} finally gives
\begin{equation*}
\curl{\lapl{\vectg}} = \emz(\partial/\partial\scalt +
\udiv{})^3\vectg.
\end{equation*}
To find the conditions under which a field described by this
equation will propagate as plane monochromatic waves of the form
\begin{equation*}
\vectg = \vectg_0\cos(\omega\scalt - \dprod{\vectkap}{\vectr})
\end{equation*}
where $\vectg_0, \omega \text{ and } \vectkap$ are independent of
time and position, let us substitute this form into the
previous equation to get the dispersion equation
\begin{equation*}
\kappa^2(\cprod{\vectkap}{\vectg}) = \emz(\omega +
\dprod{\vectu}{\vectkap})^3\vectg.
\end{equation*}
It follows from this equation that in order for the field to
propagate, \vectg\ must be parallel to $\vectkap$ and the speed
of the wave must be
\begin{equation*}
\upsilon = \omega/\kappa = -u\cos\theta
\end{equation*}
where $\theta$ is the angle between the observer's velocity
\vectu\ and the wave vector $\vectkap$ and the negative sign
indicates that the field propagates in a direction opposite to
the observer's motion. We conclude that if \eqnref{GEM:sim} holds in
a medium, then in this medium the GEM field may propagate as a
longitudinal wave with a speed that depends only on the
observer's motion and not on the material constants $\epsilon,
\mu, \zzz$ of the medium -- a sharp contrast to the familiar
behaviour of electromagnetic fields.

\section{Heat And Gravity}\label{SSPMG}
Consider again the hypothesis that \vectE\ is a function of
\vectD\ in some medium. It is easy to test this hypothesis by
experiments because according to the constitution principle,
\vectE\ must describe an electric field in the medium, and since
\vectE\ is by definition the strength of a field induced by a (time
varying) magnetic field \vectB, it follows that a time varying
magnetic field should induce an electric field in the medium.
This in turn means that under the right circumstances, the
induced electric field should produce an electric current in the
medium or cause the medium to be polarized. By similar reasoning, 
the hypothesis that \vectH\ is a
function of \vectB\ in a medium implies that a time varying
electric field should produce an emf in the medium or cause the
medium to be magnetized; and the hypothesis that \vectH\ 
(or \vectE\ ) is a function of both
\vectD\ and \vectB\ in a medium means clearly that a time varying
electric (or magnetic) field should produce both an electric
current and an emf in the medium, or cause the medium to be both
polarized and magnetized. But what about gravity ?

Well, the hypothesis that \vectH\ is a function of \vectg\ in a
medium for example implies that a time varying electric field
should produce a gravitational current in the medium. But we have
seen previously that (1) a time varying electric field
constitutes an electric current, (2) a gravitational current is a
mass flow, and (3) we know from experience that a mass flow is
almost invariably accompanied by a heat flow. We are led to
conclude therefore that if \vectH\ is a function of \vectg\ in a
medium, then an electric current should produce heat in the
medium. Conversely, if we know that an electric current produces
heat in a medium, we may argue that \vectH\ must be a function of 
\vectg\ in the medium. This line of reasoning suggests that thermoelectric
phenomena are electrogravitic in nature, for when heat produces
an electric current and an electric current produces heat in a
medium, we have in this medium that $\vectH=\vectH(\vectg),
\vectN=\vectN(\vectD)$ which together with \eqnref{GEM:Eb} and
\eqnref{GEM:Gb} describe an electrogravitic field in the medium.
Similarly, we are led to conclude that thermomagnetic phenomena
in which heat produces an emf and an emf produces heat are
essentially gravomagnetic, since in this case we have for the
medium in question that $\vectE=\vectE(\vectg),
\vectN=\vectN(\vectB)$ which together with \eqnref{GEM:Mb} and
\eqnref{GEM:Gb} describe a gravomagnetic field in the medium.

Given a medium and an arbitrary set of constitutive equations, we
are now able to verify qualitatively whether or not the medium
satisfies the equations. Hence if we have the equations
$\vectH=\vectH(\vectB), \vectE=\vectE(\vectg),
\vectN=\vectN(\vectD)$, we know that in any conducting medium 
that satisfies these equations, (a) an electric current should produce 
an emf, (b) an emf should produce heat, and (c) heat should produce an
electric current. Similarly, in any conducting medium satisfying
$\vectH=\vectH(\vectB, \vectg), \vectE=\vectE(\vectg, \vectD),
\vectN=\vectN(\vectD, \vectB)$, we have that (a) an electric
current should produce both heat and emf, (b) an emf
should produce both heat and electric current, and
(c) heat should produce both emf and electric current.
The point illustrated by these examples (which the reader may
have realized already) is that when a gravitational current
is induced in a medium in the form of heat, it indicates the
presence of an induced gravitational field in the medium in the
same way that an induced electric current indicates the presence
of an induced electric field in a medium. From this viewpoint the
results of recent experiments that suggest the existence of
gravitational shielding~\cite{Podkletnov92, Modanese96, Li97} are
by no means implausible nor inexplicable. This is more so if we
assume that matter cannot only be polarized and magnetized but
also gravitized, in which case the results of the shielding
experiments can be interpreted as indicative of the
gravitization state of matter. Further consideration of this
subject is however outside the scope of this paper.

\section{Classification Of Gravoelectromagnetic Media}\label{SSCGM}
\begin{figure}
\begin{center}
\setlength{\unitlength}{3pt}
\subfigure[$\vectH=\vectH(\vectD, \vectB,
\vectg),\vectE=\vectE(\vectD, \vectB, \vectg),
\vectN=\vectN(\vectD, \vectB, \vectg)$%
]{%
\input{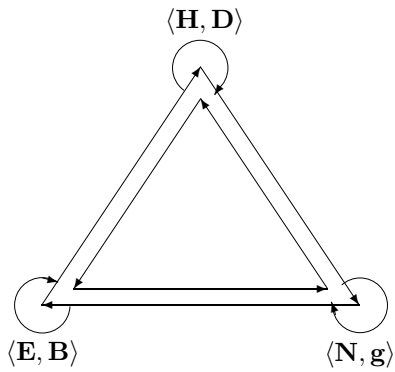}
}%
\subfigure[$\vectH=\vectH(\vectB, \vectg), \vectE=\vectE(\vectD,
\vectg ), \vectN=\vectN(\vectD, \vectB)$%
]{%
\input{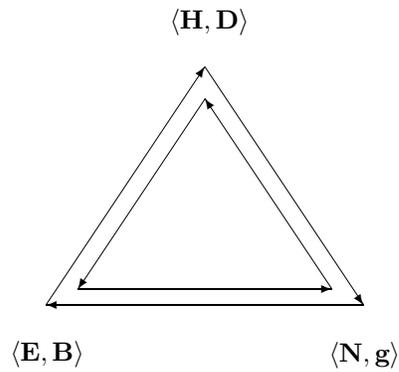}
}\\%
\subfigure[$\vectH=\vectH(\vectD, \vectg), \vectE=\vectE(\vectD),
\vectN=\vectN(\vectD, \vectB)$%
]{%
\input{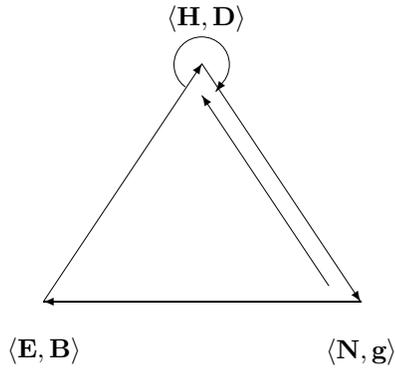}\label{FIG1C}
}%
\subfigure[$\vectH=\vectH(\vectD), \vectE=\vectE(\vectB),
\vectN=\vectN(\vectg)$%
]{%
\input{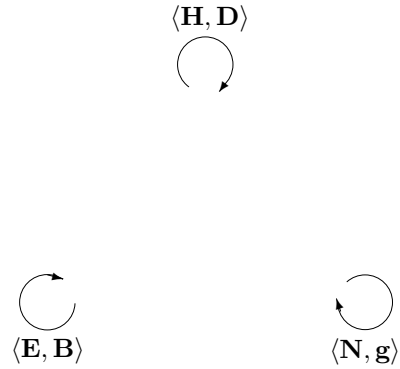}
}%
\caption{\footnotesize Examples of constitutive diagrams showing (a) all
possible interactions (b) no self interactions (c) mixed
interactions (d) no mutual interactions} \label{FIG1}
\end{center}
\end{figure}
It is often instructive to show the structure of a set of
constitutive equations in a so-called constitutive diagram. To do
so, we write down the pairs \brackD, \brackB\ and \brackg. Then,
if \vectH\ is a function of \vectB, for example, we draw an arrow
from the pair containing \vectH\ to the pair containing \vectB;
that is, from \brackD\ to \brackB. As the examples in
\figref{FIG1} show, constitutive diagrams make it easy to see the
interactions described by a set of constitutive equations. They
also make it easy to see the types of unified fields implied by
these interactions.
Thus in \figref{FIG1C}, the arrows \brackD\induces\brackg,
\brackg\induces\brackB, and \brackB\induces\brackD\ indicate a
gravoelectromagnetic field because they form a closed loop
containing all three fields. Since the arrows
\brackD\induces\brackg\ and \brackg\induces\brackD\ also form a
closed loop indicating an electrogravitic field, it follows that
any medium whose constitutive equations correspond to
\figref{FIG1C} will be able to sustain both a
gravoelectromagnetic field and an electrogravitic field.

Considering that a constitutive equation such as
$\vectH=\vectH(\vectD, \vectB)$ describes a class of media rather than a
single medium, it is useful to identify all classes of media
described in this way together with the unified fields that each
class supports. To perform this exercise we shall first introduce
the following rules for writing constitutive equations in a
shorthand form~: (1) omit the equals sign and everything to its
left; (2) omit the opening and closing parentheses on the right
hand side; (3) omit any commas on the right hand side; (4) write
the vectors as scalars; (5) write the constitutive equations for
\vectH, \vectE, \vectN\ in that order, separated by hyphens and
subject to the preceding conventions. By these rules, for
examples, the constitutive equations in \figref{FIG1C} will become
\hbox{HDg-ED-NDB} while \eqnref{G:cona} will become HB-Eg-ND and
\eqnref{G:conb} will become Hg-ED-NB. \tabref{TABA} shows the result
of the exercise. We refer to the columns (numbered 1 to 7) as
groups, to the rows (lettered A to G) as clusters, and to the
nested rows (lettered a to g) as families. To define a set of
constitutive equations with reference to this table, one needs to
give the relevant cluster, group and family in that order.

For example, the constitutive equations $\vectH=\vectH(\vectD,
\vectB), \vectE=\vectE(\vectB, \vectg), \vectN=\vectN(\vectD,
\vectg)$ yield the shorthand equation (also called the signature)
\signat{DB}{Bg}{Dg} which according to \tabref{TABA} describes a
class D6e medium. Conversely, for a class B3a medium,
\tabref{TABA} gives the signature \signat{B}{g}{D} which in turn
gives the constitutive equations $\vectH=\vectH(\vectB),
\vectE=\vectE(\vectg), \vectN=\vectN(\vectD)$.
The constitution table is useful because it
categorizes media with similar properties together in the
same way that the periodic table categorizes the chemical elements. 
Group 3 media
for example have the property that in them, \vectE\ is a function
of \vectg, which means that in these media an emf can produce
heat. Similarly, cluster F media can be seen to have the property
that in them \vectH\ is a function of \vectB\ and \vectg, which
means that in these media an electric current can produce both
heat and emf.

\begin{table}
\begin{narrow}{-23mm}{-8mm}
\myf
\begin{tabular}{|l|l|l|l|l|l|l|l|l|}
\hline
  \multicolumn{2}{|c}{}
    & \multicolumn{1}{|c}{1} & \ctr{2} & \ctr{3} & \ctr{4} & \ctr{5} & \ctr{6} & \multicolumn{1}{c|}{7}\\
\hline
   & a. & \sig{D}{D}{D}{\xnnnn}     & \sig{D}{B}{D}{\xnnnn}     & \sig{D}{g}{D}{\xnnnn}     & \sig{D}{DB}{D}{\xnnnn}    & \sig{D}{Dg}{D}{\xnnnn}    & \sig{D}{Bg}{D}{\xnnnn}    & \sig{D}{DBg}{D}{\xnnnn}   \\
   & b. & \sig{D}{D}{B}{\xnnnn}     & \sig{D}{B}{B}{\xnnnn}     & \sig{D}{g}{B}{\xcnnn}     & \sig{D}{DB}{B}{\xnnnn}    & \sig{D}{Dg}{B}{\xcnnn}    & \sig{D}{Bg}{B}{\xcnnn}    & \sig{D}{DBg}{B}{\xcnnn}   \\
   & c. & \sig{D}{D}{g}{\xnnnn}     & \sig{D}{B}{g}{\xnnnn}     & \sig{D}{g}{g}{\xnnnn}     & \sig{D}{DB}{g}{\xnnnn}    & \sig{D}{Dg}{g}{\xnnnn}    & \sig{D}{Bg}{g}{\xnnnn}    & \sig{D}{DBg}{g}{\xnnnn}   \\
A  & d. & \sig{D}{D}{DB}{\xnnnn}    & \sig{D}{B}{DB}{\xnnnn}    & \sig{D}{g}{DB}{\xcnnn}    & \sig{D}{DB}{DB}{\xnnnn}   & \sig{D}{Dg}{DB}{\xcnnn}   & \sig{D}{Bg}{DB}{\xcnnn}   & \sig{D}{DBg}{DB}{\xcnnn}  \\
   & e. & \sig{D}{D}{Dg}{\xnnnn}    & \sig{D}{B}{Dg}{\xnnnn}    & \sig{D}{g}{Dg}{\xnnnn}    & \sig{D}{DB}{Dg}{\xnnnn}   & \sig{D}{Dg}{Dg}{\xnnnn}   & \sig{D}{Bg}{Dg}{\xnnnn}   & \sig{D}{DBg}{Dg}{\xnnnn}  \\
   & f. & \sig{D}{D}{Bg}{\xnnnn}    & \sig{D}{B}{Bg}{\xnnnn}    & \sig{D}{g}{Bg}{\xcnnn}    & \sig{D}{DB}{Bg}{\xnnnn}   & \sig{D}{Dg}{Bg}{\xcnnn}   & \sig{D}{Bg}{Bg}{\xcnnn}   & \sig{D}{DBg}{Bg}{\xcnnn}  \\
   & g. & \sig{D}{D}{DBg}{\xnnnn}   & \sig{D}{B}{DBg}{\xnnnn}   & \sig{D}{g}{DBg}{\xcnnn}   & \sig{D}{DB}{DBg}{\xnnnn}  & \sig{D}{Dg}{DBg}{\xcnnn}  & \sig{D}{Bg}{DBg}{\xcnnn}  & \sig{D}{DBg}{DBg}{\xcnnn} \\
\cline{1-9}
   & a. & \sig{B}{D}{D}{\xannn}     & \sig{B}{B}{D}{\xnnnn}     & \sig{B}{g}{D}{\xdnnn}     & \sig{B}{DB}{D}{\xannn}    & \sig{B}{Dg}{D}{\xadnn}    & \sig{B}{Bg}{D}{\xdnnn}    & \sig{B}{DBg}{D}{\xadnn}   \\
   & b. & \sig{B}{D}{B}{\xannn}     & \sig{B}{B}{B}{\xnnnn}     & \sig{B}{g}{B}{\xcnnn}     & \sig{B}{DB}{B}{\xannn}    & \sig{B}{Dg}{B}{\xacdn}    & \sig{B}{Bg}{B}{\xcnnn}    & \sig{B}{DBg}{B}{\xacdn}   \\
   & c. & \sig{B}{D}{g}{\xannn}     & \sig{B}{B}{g}{\xnnnn}     & \sig{B}{g}{g}{\xnnnn}     & \sig{B}{DB}{g}{\xannn}    & \sig{B}{Dg}{g}{\xannn}    & \sig{B}{Bg}{g}{\xnnnn}    & \sig{B}{DBg}{g}{\xannn}   \\
B  & d. & \sig{B}{D}{DB}{\xannn}    & \sig{B}{B}{DB}{\xnnnn}    & \sig{B}{g}{DB}{\xcdnn}    & \sig{B}{DB}{DB}{\xannn}   & \sig{B}{Dg}{DB}{\xacdn}   & \sig{B}{Bg}{DB}{\xcdnn}   & \sig{B}{DBg}{DB}{\xacdn}  \\
   & e. & \sig{B}{D}{Dg}{\xannn}    & \sig{B}{B}{Dg}{\xnnnn}    & \sig{B}{g}{Dg}{\xdnnn}    & \sig{B}{DB}{Dg}{\xannn}   & \sig{B}{Dg}{Dg}{\xadnn}   & \sig{B}{Bg}{Dg}{\xdnnn}   & \sig{B}{DBg}{Dg}{\xadnn}  \\
   & f. & \sig{B}{D}{Bg}{\xannn}    & \sig{B}{B}{Bg}{\xnnnn}    & \sig{B}{g}{Bg}{\xcnnn}    & \sig{B}{DB}{Bg}{\xannn}   & \sig{B}{Dg}{Bg}{\xacdn}   & \sig{B}{Bg}{Bg}{\xcnnn}   & \sig{B}{DBg}{Bg}{\xacdn}  \\
   & g. & \sig{B}{D}{DBg}{\xannn}   & \sig{B}{B}{DBg}{\xnnnn}   & \sig{B}{g}{DBg}{\xcdnn}   & \sig{B}{DB}{DBg}{\xannn}  & \sig{B}{Dg}{DBg}{\xacdn}  & \sig{B}{Bg}{DBg}{\xcdnn}  & \sig{B}{DBg}{DBg}{\xacdn} \\
\cline{1-9}
   & a. & \sig{g}{D}{D}{\xbnnn}     & \sig{g}{B}{D}{\xbnnn}     & \sig{g}{g}{D}{\xbnnn}     & \sig{g}{DB}{D}{\xbnnn}    & \sig{g}{Dg}{D}{\xbnnn}    & \sig{g}{Bg}{D}{\xbnnn}    & \sig{g}{DBg}{D}{\xbnnn}   \\
   & b. & \sig{g}{D}{B}{\xdnnn}     & \sig{g}{B}{B}{\xnnnn}     & \sig{g}{g}{B}{\xcnnn}     & \sig{g}{DB}{B}{\xdnnn}    & \sig{g}{Dg}{B}{\xcdnn}    & \sig{g}{Bg}{B}{\xcnnn}    & \sig{g}{DBg}{B}{\xcdnn}   \\
   & c. & \sig{g}{D}{g}{\xnnnn}     & \sig{g}{B}{g}{\xnnnn}     & \sig{g}{g}{g}{\xnnnn}     & \sig{g}{DB}{g}{\xnnnn}    & \sig{g}{Dg}{g}{\xnnnn}    & \sig{g}{Bg}{g}{\xnnnn}    & \sig{g}{DBg}{g}{\xnnnn}   \\
C  & d. & \sig{g}{D}{DB}{\xbdnn}    & \sig{g}{B}{DB}{\xbnnn}    & \sig{g}{g}{DB}{\xbcdn}    & \sig{g}{DB}{DB}{\xbdnn}   & \sig{g}{Dg}{DB}{\xbcdn}   & \sig{g}{Bg}{DB}{\xbcdn}   & \sig{g}{DBg}{DB}{\xbcdn}  \\
   & e. & \sig{g}{D}{Dg}{\xbnnn}    & \sig{g}{B}{Dg}{\xbnnn}    & \sig{g}{g}{Dg}{\xbnnn}    & \sig{g}{DB}{Dg}{\xbnnn}   & \sig{g}{Dg}{Dg}{\xbnnn}   & \sig{g}{Bg}{Dg}{\xbnnn}   & \sig{g}{DBg}{Dg}{\xbnnn}  \\
   & f. & \sig{g}{D}{Bg}{\xdnnn}    & \sig{g}{B}{Bg}{\xnnnn}    & \sig{g}{g}{Bg}{\xcnnn}    & \sig{g}{DB}{Bg}{\xdnnn}   & \sig{g}{Dg}{Bg}{\xcdnn}   & \sig{g}{Bg}{Bg}{\xcnnn}   & \sig{g}{DBg}{Bg}{\xcdnn}  \\
   & g. & \sig{g}{D}{DBg}{\xbdnn}   & \sig{g}{B}{DBg}{\xbnnn}   & \sig{g}{g}{DBg}{\xbcdn}   & \sig{g}{DB}{DBg}{\xbdnn}  & \sig{g}{Dg}{DBg}{\xbcdn}  & \sig{g}{Bg}{DBg}{\xbcdn}  & \sig{g}{DBg}{DBg}{\xbcdn} \\
\cline{1-9}
   & a. & \sig{DB}{D}{D}{\xannn}     & \sig{DB}{B}{D}{\xnnnn}     & \sig{DB}{g}{D}{\xdnnn}     & \sig{DB}{DB}{D}{\xannn}    & \sig{DB}{Dg}{D}{\xadnn}    & \sig{DB}{Bg}{D}{\xdnnn}    & \sig{DB}{DBg}{D}{\xadnn}   \\
   & b. & \sig{DB}{D}{B}{\xannn}     & \sig{DB}{B}{B}{\xnnnn}     & \sig{DB}{g}{B}{\xcnnn}     & \sig{DB}{DB}{B}{\xannn}    & \sig{DB}{Dg}{B}{\xacdn}    & \sig{DB}{Bg}{B}{\xcnnn}    & \sig{DB}{DBg}{B}{\xacdn}   \\
   & c. & \sig{DB}{D}{g}{\xannn}     & \sig{DB}{B}{g}{\xnnnn}     & \sig{DB}{g}{g}{\xnnnn}     & \sig{DB}{DB}{g}{\xannn}    & \sig{DB}{Dg}{g}{\xannn}    & \sig{DB}{Bg}{g}{\xnnnn}    & \sig{DB}{DBg}{g}{\xannn}   \\
D  & d. & \sig{DB}{D}{DB}{\xannn}    & \sig{DB}{B}{DB}{\xnnnn}    & \sig{DB}{g}{DB}{\xcdnn}    & \sig{DB}{DB}{DB}{\xannn}   & \sig{DB}{Dg}{DB}{\xacdn}   & \sig{DB}{Bg}{DB}{\xcdnn}   & \sig{DB}{DBg}{DB}{\xacdn}  \\
   & e. & \sig{DB}{D}{Dg}{\xannn}    & \sig{DB}{B}{Dg}{\xnnnn}    & \sig{DB}{g}{Dg}{\xdnnn}    & \sig{DB}{DB}{Dg}{\xannn}   & \sig{DB}{Dg}{Dg}{\xadnn}   & \sig{DB}{Bg}{Dg}{\xcdnn}   & \sig{DB}{DBg}{Dg}{\xadnn}  \\
   & f. & \sig{DB}{D}{Bg}{\xannn}    & \sig{DB}{B}{Bg}{\xnnnn}    & \sig{DB}{g}{Bg}{\xcnnn}    & \sig{DB}{DB}{Bg}{\xannn}   & \sig{DB}{Dg}{Bg}{\xacdn}   & \sig{DB}{Bg}{Bg}{\xcnnn}   & \sig{DB}{DBg}{Bg}{\xacdn}  \\
   & g. & \sig{DB}{D}{DBg}{\xannn}   & \sig{DB}{B}{DBg}{\xnnnn}   & \sig{DB}{g}{DBg}{\xcdnn}   & \sig{DB}{DB}{DBg}{\xannn}  & \sig{DB}{Dg}{DBg}{\xacdn}  & \sig{DB}{Bg}{DBg}{\xcdnn}  & \sig{DB}{DBg}{DBg}{\xacdn} \\
\cline{1-9}
   & a. & \sig{Dg}{D}{D}{\xbnnn}     & \sig{Dg}{B}{D}{\xbnnn}     & \sig{Dg}{g}{D}{\xbnnn}     & \sig{Dg}{DB}{D}{\xbnnn}    & \sig{Dg}{Dg}{D}{\xbnnn}    & \sig{Dg}{Bg}{D}{\xbnnn}    & \sig{Dg}{DBg}{D}{\xbnnn}   \\
   & b. & \sig{Dg}{D}{B}{\xdnnn}     & \sig{Dg}{B}{B}{\xnnnn}     & \sig{Dg}{g}{B}{\xcnnn}     & \sig{Dg}{DB}{B}{\xdnnn}    & \sig{Dg}{Dg}{B}{\xcdnn}    & \sig{Dg}{Bg}{B}{\xcnnn}    & \sig{Dg}{DBg}{B}{\xcdnn}   \\
   & c. & \sig{Dg}{D}{g}{\xnnnn}     & \sig{Dg}{B}{g}{\xnnnn}     & \sig{Dg}{g}{g}{\xnnnn}     & \sig{Dg}{DB}{g}{\xnnnn}    & \sig{Dg}{Dg}{g}{\xnnnn}    & \sig{Dg}{Bg}{g}{\xnnnn}    & \sig{Dg}{DBg}{g}{\xnnnn}   \\
E  & d. & \sig{Dg}{D}{DB}{\xbdnn}    & \sig{Dg}{B}{DB}{\xbnnn}    & \sig{Dg}{g}{DB}{\xbcdn}    & \sig{Dg}{DB}{DB}{\xbdnn}   & \sig{Dg}{Dg}{DB}{\xbcdn}   & \sig{Dg}{Bg}{DB}{\xbcdn}   & \sig{Dg}{DBg}{DB}{\xbcdn}  \\
   & e. & \sig{Dg}{D}{Dg}{\xbnnn}    & \sig{Dg}{B}{Dg}{\xbnnn}    & \sig{Dg}{g}{Dg}{\xbnnn}    & \sig{Dg}{DB}{Dg}{\xbnnn}   & \sig{Dg}{Dg}{Dg}{\xbnnn}   & \sig{Dg}{Bg}{Dg}{\xbnnn}   & \sig{Dg}{DBg}{Dg}{\xbnnn}  \\
   & f. & \sig{Dg}{D}{Bg}{\xdnnn}    & \sig{Dg}{B}{Bg}{\xnnnn}    & \sig{Dg}{g}{Bg}{\xcnnn}    & \sig{Dg}{DB}{Bg}{\xdnnn}   & \sig{Dg}{Dg}{Bg}{\xcdnn}   & \sig{Dg}{Bg}{Bg}{\xcnnn}   & \sig{Dg}{DBg}{Bg}{\xcdnn}  \\
   & g. & \sig{Dg}{D}{DBg}{\xbdnn}   & \sig{Dg}{B}{DBg}{\xbnnn}   & \sig{Dg}{g}{DBg}{\xbcdn}   & \sig{Dg}{DB}{DBg}{\xbdnn}  & \sig{Dg}{Dg}{DBg}{\xbcdn}  & \sig{Dg}{Bg}{DBg}{\xbcdn}  & \sig{Dg}{DBg}{DBg}{\xbcdn} \\
\cline{1-9}
   & a. & \sig{Bg}{D}{D}{\xabdn}     & \sig{Bg}{B}{D}{\xbnnn}     & \sig{Bg}{g}{D}{\xbdnn}     & \sig{Bg}{DB}{D}{\xabdn}    & \sig{Bg}{Dg}{D}{\xabdn}    & \sig{Bg}{Bg}{D}{\xbdnn}    & \sig{Bg}{DBg}{D}{\xabdn}   \\
   & b. & \sig{Bg}{D}{B}{\xadnn}     & \sig{Bg}{B}{B}{\xnnnn}     & \sig{Bg}{g}{B}{\xcnnn}     & \sig{Bg}{DB}{B}{\xadnn}    & \sig{Bg}{Dg}{B}{\xacdn}    & \sig{Bg}{Bg}{B}{\xcnnn}    & \sig{Bg}{DBg}{B}{\xacdn}   \\
   & c. & \sig{Bg}{D}{g}{\xannn}     & \sig{Bg}{B}{g}{\xnnnn}     & \sig{Bg}{g}{g}{\xnnnn}     & \sig{Bg}{DB}{g}{\xannn}    & \sig{Bg}{Dg}{g}{\xannn}    & \sig{Bg}{Bg}{g}{\xnnnn}    & \sig{Bg}{DBg}{g}{\xannn}   \\
F  & d. & \sig{Bg}{D}{DB}{\xabdn}    & \sig{Bg}{B}{DB}{\xbnnn}    & \sig{Bg}{g}{DB}{\xbcdn}    & \sig{Bg}{DB}{DB}{\xabdn}   & \sig{Bg}{Dg}{DB}{\xabcd}   & \sig{Bg}{Bg}{DB}{\xbcdn}   & \sig{Bg}{DBg}{DB}{\xabcd}  \\
   & e. & \sig{Bg}{D}{Dg}{\xabdn}    & \sig{Bg}{B}{Dg}{\xbnnn}    & \sig{Bg}{g}{Dg}{\xbdnn}    & \sig{Bg}{DB}{Dg}{\xabdn}   & \sig{Bg}{Dg}{Dg}{\xabdn}   & \sig{Bg}{Bg}{Dg}{\xbdnn}   & \sig{Bg}{DBg}{Dg}{\xabdn}  \\
   & f. & \sig{Bg}{D}{Bg}{\xadnn}    & \sig{Bg}{B}{Bg}{\xnnnn}    & \sig{Bg}{g}{Bg}{\xcnnn}    & \sig{Bg}{DB}{Bg}{\xadnn}   & \sig{Bg}{Dg}{Bg}{\xacdn}   & \sig{Bg}{Bg}{Bg}{\xcnnn}   & \sig{Bg}{DBg}{Bg}{\xacdn}  \\
   & g. & \sig{Bg}{D}{DBg}{\xabdn}   & \sig{Bg}{B}{DBg}{\xbnnn}   & \sig{Bg}{g}{DBg}{\xbcdn}   & \sig{Bg}{DB}{DBg}{\xabdn}  & \sig{Bg}{Dg}{DBg}{\xabcd}  & \sig{Bg}{Bg}{DBg}{\xbcdn}  & \sig{Bg}{DBg}{DBg}{\xabcd} \\
\cline{1-9}
   & a. & \sig{DBg}{D}{D}{\xabdn}     & \sig{DBg}{B}{D}{\xbnnn}     & \sig{DBg}{g}{D}{\xbdnn}     & \sig{DBg}{DB}{D}{\xabdn}    & \sig{DBg}{Dg}{D}{\xabdn}    & \sig{DBg}{Bg}{D}{\xbdnn}    & \sig{DBg}{DBg}{D}{\xabdn}   \\
   & b. & \sig{DBg}{D}{B}{\xadnn}     & \sig{DBg}{B}{B}{\xnnnn}     & \sig{DBg}{g}{B}{\xcnnn}     & \sig{DBg}{DB}{B}{\xadnn}    & \sig{DBg}{Dg}{B}{\xacdn}    & \sig{DBg}{Bg}{B}{\xcnnn}    & \sig{DBg}{DBg}{B}{\xacdn}   \\
   & c. & \sig{DBg}{D}{g}{\xannn}     & \sig{DBg}{B}{g}{\xnnnn}     & \sig{DBg}{g}{g}{\xnnnn}     & \sig{DBg}{DB}{g}{\xannn}    & \sig{DBg}{Dg}{g}{\xannn}    & \sig{DBg}{Bg}{g}{\xnnnn}    & \sig{DBg}{DBg}{g}{\xannn}   \\
G  & d. & \sig{DBg}{D}{DB}{\xabdn}    & \sig{DBg}{B}{DB}{\xbnnn}    & \sig{DBg}{g}{DB}{\xbcdn}    & \sig{DBg}{DB}{DB}{\xabdn}   & \sig{DBg}{Dg}{DB}{\xabcd}   & \sig{DBg}{Bg}{DB}{\xbcdn}   & \sig{DBg}{DBg}{DB}{\xabcd}  \\
   & e. & \sig{DBg}{D}{Dg}{\xabdn}    & \sig{DBg}{B}{Dg}{\xbnnn}    & \sig{DBg}{g}{Dg}{\xbdnn}    & \sig{DBg}{DB}{Dg}{\xabdn}   & \sig{DBg}{Dg}{Dg}{\xabdn}   & \sig{DBg}{Bg}{Dg}{\xbdnn}   & \sig{DBg}{DBg}{Dg}{\xabdn}  \\
   & f. & \sig{DBg}{D}{Bg}{\xadnn}    & \sig{DBg}{B}{Bg}{\xnnnn}    & \sig{DBg}{g}{Bg}{\xcnnn}    & \sig{DBg}{DB}{Bg}{\xadnn}   & \sig{DBg}{Dg}{Bg}{\xacdn}   & \sig{DBg}{Bg}{Bg}{\xcnnn}   & \sig{DBg}{DBg}{Bg}{\xacdn}  \\
   & g. & \sig{DBg}{D}{DBg}{\xabdn}   & \sig{DBg}{B}{DBg}{\xbnnn}   & \sig{DBg}{g}{DBg}{\xbcdn}   & \sig{DBg}{DB}{DBg}{\xabdn}  & \sig{DBg}{Dg}{DBg}{\xabcd}  & \sig{DBg}{Bg}{DBg}{\xbcdn}  & \sig{DBg}{DBg}{DBg}{\xabcd} \\
\hline
\end{tabular}
\caption{\footnotesize Constitution table showing all classes of
gravoelectromagnetic media. The unified field supported by
each class is indicated as \embox\ electromagnetic, \gebox\
electrogravitic, \gmbox\ gravomagnetic, \gembox\
gravoelectromagnetic, \nbox\ none. \copyright1996, 2001 Adequest
International Corp. } 
\label{TABA}
\end{narrow}
\end{table}


\section{Discussion And Conclusion}\label{SSDAC}
Equations \eqnref{GEM:E} and \eqnref{GEM:M} are the same equations
postulated by Hertz in 1892 except that Hertz added another
parameter to \eqnref{GEM:Ec} and assumed that \vectu\ described the
absolute motion of an observer relative to a luminiferous ether.
The additional parameter was thought to be necessary in order for
the theory to treat convection and conduction currents; it has
been omitted from consideration because one can treat these
currents via \vectj\ as \eqnref{E:j} shows. 
If one is not interested in the motion that produces \vectj, then one
can ignore this equation and treat \vectj\ as in Maxwell's
theory. If one is interested in this motion,
however, then one must take \eqnref{E:j} into account. In this
and all other equations vector \vectu\ is the
velocity of point \vectr\ where one is observing a field at time
\scalt\ and can be interpreted as the velocity of an observer 
(or field detector, probe, etc) at
this point; it is not measured relative to ether but relative to
the point that one chooses as the origin of coordinates for the
position vector \vectr\ in the analytic expressions for the field
vectors \vectD(\vectr, \scalt), \vectB(\vectr, \scalt) and
\vectg(\vectr, \scalt). If $\delta = 0, \vectb = \vect{0}$ and
one ignores both \eqnref{GEM:Ec} and \eqnref{GEM:Mc}, then \eqnref{GEM:E}
and \eqnref{GEM:M} become Maxwell's equations. This means that
Hertz's theory satisfies the correspondence principle in the
sense that all phenomena described by Maxwell's equations are
also described by Hertz's equations. Since Maxwell's theory
describes the electrodynamics of stationary bodies (as is well
known), formal correspondence exists with Hertz's theory only for
phenomena related to the electrodynamics of stationary bodies.

For phenomena related to the electrodynamics of moving bodies,
significant differences between both theories exist. These
differences arise from the fact that problems of moving bodies
electrodynamics must be treated in Maxwell's theory via special
relativity and in Hertz's theory via vector \vectu. Equation
\eqnref{GEM:Ec} for example shows how the current density \vectj\
arises from motion in Hertz's theory. While Maxwell's theory also
recognizes that \vectj\ arises from motion, the theory gives no
general expression for \vectj\ in terms of this motion.
Historically, this situation arose because when Maxwell first
introduced \eqnref{E:di} into his theory, he considered $\vectJE$ to
be an independent parameter. He later realized that for a time
varying field, $\vectJE$ could not strictly be independent of
\vectD. So he introduced the displacement density
$\partial\vectD/\partial\scalt$ into his theory and by doing this
made $\vectJE$ depend partially on \vectD\ while retaining
\vectj\ as an independent parameter~\cite{Buchwald85, Roche98}.
Hertz would argue that since
the displacement density arises from temporal changes in \vectD,
it is natural to suppose that the total density $\vectJE$ also
does. This is why $\vectJE$ is defined in Hertz's theory as the
total time derivative of \vectD. The corresponding definition for
$\vectJM$ follows by symmetry and Lenz's law.

Hertz's theory suggests that one should call \vectE\ the strength
or induction of a magnetoinduced (i.e. magnetically induced)
field and \vectH\ the strength or 
induction of an electroinduced (i.e. electrically induced)
field because by the constitution principle, the nature of the fields
described by these vectors depend on the constitutive equations
of a medium and, for this reason, one should not describe
\vectE\ as electric nor \vectH\ as magnetic without further
qualifications. Also, the theory
implies that an electric current or emf may exist in situations
where neither would be expected according to Maxwell's theory.
Equation \eqnref{M:xib} shows for example that an emf may exist when
a magnetic flux is not changing with time ($d\Phi/d\scalt = 0$),
contrary to \eqnref{M:mwl1}. We know that an emf has been observed
in the Kennard-M\"uller homopolar induction effect that
apparently should not have existed according to Maxwell's theory
\cite{Kennard17, Muller90}. This gives qualitative support for
\eqnref{M:xib}; quantitative support can be obtained by showing that
the observed emf in this effect is consistent with \eqnref{M:xib},
but this is outside the scope of this paper. Interestingly,
according to \eqnref{M:ba}, the extraneous emf due to \vectb\ should
exist for $\delta = 0$ as well as for $\delta\ne0$. This means
that one cannot omit \vectb\ from \eqnref{GEM:Mb} on the grounds
that $\delta = 0$ identically.

It is also interesting to note that Ampere's law and Faraday's
law are valid only when $di/dV = 0$ and $d\xi/dV = 0$ in a
region. From \eqnref{E:divJE} and \eqnref{E:JE}, we get
\begin{equation*}
\frac{di}{dV} = \frac{\partial\sigma}{\partial\scalt} +
\dive{\vectj}
\end{equation*}
while from \eqnref{M:divJM} and \eqnref{M:JM}, we get
\begin{equation*}
\frac{d\xi}{dV} = -\frac{\partial\delta}{\partial\scalt} -
\dive{\vectb}
\end{equation*}
both of which will be recognized as continuity equations for
\vectD\ and \vectB. This means that Ampere's law and Faraday's law
are valid only when there is conservation of electric charge and
magnetic flux in a region; or, equivalently, when the electric
current \scali\ and the emf $\xi$ in a region are both
independent of the volume of the region. This is known to be the
case from elementary experiments on thin and thick wires on the
one hand, and from similar experiments on small and large voltaic
cells on the other.

Regarding gravity, we have shown that the introduction of
parameter \scalh\ invariably leads to the existence of vector
\vectN\ which is the gravitational analogue of \vectE\ and \vectH.
Using this vector and the constitution principle, we gave a clear
picture of how electric, magnetic, and gravitational fields can
interact in a medium to form more complex fields. Among other
things our method reduces the problem of unifying
electromagnetism and gravity to the problem of finding the
precise form of \eqnref{GEM:con} for a medium. The latter problem
can be treated by studying the relationship between the field
inductions \vectH, \vectE, \vectN\ and the field intensities
\vectD, \vectB, \vectg\ for different media experimentally using
the hypothesis that a gravitational current manifests as heat, or
by postulating different forms of \eqnref{GEM:con} and studying the
resulting properties of the unified field theoretically. Both
approaches have led to interesting results which are outside the scope 
of this paper.
There are of course several ways in which the method of unification
described here can be further developed. These include a theory
of the constitution table explaining the formation of each class
of GEM media, a retarded integral formulation of the GEM field
equations similarly to Jefimenko's theory~\cite{Jefimenko00}, and
a more detailed investigation of the relationship between heat and
gravity. In addition, as many useful solutions to the problems 
of moving bodies electrodynamics as are possible need to be obtained
in order to realize the full potential of the unification method described
in this paper.

\subsection*{Dedication}
This paper is dedicated to the late Mr. Sunday Otusanya, of 
loving and blessed memory, with a prayer that his soul rests in peace, amen.

\bibliographystyle{unsrt}
\bibliography{xbib}

\section*{Errata 21may2001}
\begin{itemize}
\item rename ``principle of induction'' to ``principle of constitution''
\item minor corrections in spelling, grammar and symbols
\item changes in format and copyright notice
\end{itemize}

\end{document}